\documentclass{aa}

\newcommand{\xmm}{{\it XMM-Newton}}

\newcommand{\kepler}{{\it Kepler}}

\newcommand{\ginga}{{\it Ginga}}

\newcommand{\astrosat}{{\it AstroSat}}
\newcommand{\nicer}{{\it NICER}}
\newcommand{\ixpe}{{\it IXPE}}

\begin{document}

\title{Searching for similarities in the accretion flow of Seyfert 1 galaxies and cataclysmic variables based on flare profiles of IRAS\,13224-3809, 1H\,0707-495, Mrk\,766 and MV\,Lyr}
\titlerunning{Flare profiles in AGNs and CVs}
\authorrunning{Dobrotka et al.}

\author{A.~Dobrotka \inst {1}, H.~Negoro \inst {2} and P.~Bez\'ak \inst {1}}

\offprints{A.~Dobrotka, \email{andrej.dobrotka@stuba.sk}}

\institute{Advanced Technologies Research Institute, Faculty of Materials Science and Technology in Trnava, Slovak University of Technology in Bratislava, Bottova 25, 917 24 Trnava, Slovakia
\and
Department of Physics, Nihon University, 1-8 Kanda-Surugadai, Chiyoda-ku, Tokyo 101-8308, Japan
}

\date{Received / Accepted}

\abstract
{}
{We studied fast variability of three selected AGNs, IRAS\,13224-3809, 1H\,0707-495 and Mrk\,766, and the cataclysmic variable MV\,Lyr observed by \xmm\ and \kepler\ spacecrafts, respectively. Our goal is to search for common origin of the variability and to test the so-called sandwich model where a geometrically thick corona is surrounding a geometrically thin disc.}
{We study substructures of the averaged flare profiles. The flare profile method identifies individual flares in the light curve, and averages them. Direct fitting of the profile substructures identify individual characteristic frequencies seen in standard power density spectra (PDS) as a break frequency or quasi-periodic oscillation. The credibility of flare profile substructures is demonstrated by comparison with autocorrelation function.}
{We found that the flare profiles of AGNs are similar to that of the cataclysmic variable in the low state. We explain this as a consequence of a truncated inner disc in a sandwich model. The same scenario is also able to explain the presence of characteristic break frequencies in X-ray PDS, but not seen in optical. We also searched for substructures in the flare profile of IRAS\,13224-3809. In addition to a permanently present main flare, we found transient side-lobe appearing before the main flare and only seen in a high flux period. This complex flare profile of this AGN suggests that an additional source of X-rays appears during the high flux period. We propose a scenario in which an accretion flow fluctuation enters the sandwich corona and propagates further to some very central part of the accretion disc.}
{}


\maketitle

\section{Introduction}
\label{introduction}

A large number of objects, such as cataclysmic variables (CVs), X-ray binaries (XRBs) or active galactic nuclei (AGNs), are driven by a common physical process: accretion. The gas in these systems falls towards the central compact object and, in the absence of a strong magnetic field, an accretion disk forms. The central accretor may be a white dwarf (WD) in the case of a CV, or a neutron star or stellar black hole (BH) for a XRB and a supermassive BH in an AGN. An accretion disc has many potential substructures and forms depending on the central object character. CVs systems possess a boundary layer at the interaction between the inner disc and the WD. Such a boundary layer is missing in the case of BHs in XRBs or AGNs. An inner geometrically thin disc can be evaporated and a hot geometrically thick corona can be formed. The diversity is large, but the basic assumption is that if every mentioned accretion system is powered by the same physical process, they must have similar manifestations.

The most typical radiation pattern of the accreting process is in the form of stochastic variability called flickering. While the underlying emission mechanisms are relatively well understood, the location and exact shape of individual sources are still unclear.

AGNs exhibit the strongest and fastest variability in X-rays on various time-scales (see e.g. \citealt{padovani2017} for a review). These time-scales are as short as $\sim 100$\,s (e.g. \citealt{vaughan2011}). This implies a very small emission region located at the inner accretion disc. The longest time-scales of these objects are up to several days (see e.g. \citealt{edelson2014}) or more, while CVs reaches only hours (see e.g. \citealt{scaringi2012a}). However, the shortest time-scales are comparable in both objects (see e.g. \citealt{dobrotka2017a,dobrotka2019} for CVs).

Every physical mechanism generating such variability has its characteristic time-scale or range of time-scales. The most popular way how to measure such a time-scale is the study using power density spectra (PDS). It has the shape of a red noise or a band limited noise with break frequencies or Lorentzian components indicating presence of QPOs (\citealt{scaringi2012a,dobrotka2016,alston2019}).

The PDS shape with other properties like linear correlation between variability amplitude and log-normally distributed flux (so called rms-flux relation) observed in all variety of accreting systems such as XRBs (\citealt{negoro2002}), AGNs (\citealt{uttley2005}), or CVs (\citealt{scaringi2012b,zamanov2015}) is well explained by accretion fluctuation propagation scenario (\citealt{manmoto1996,lyubarskii1997,kotov2001,arevalo2006}). Following this model every accretion rate fluctuation generated anywhere in the disc is propagating inside,
and gives rise to radiation by releasing gravitational energy. As a result all characteristic time-scale and processes are encoded in the observed light curve.

An alternative method to study characteristic time-scales seen in the PDS was proposed by \citet{negoro1994}. The authors superposed many flares\footnote{Original work by \citet{negoro1994} uses word "shot".} from \ginga\ observations of the XRB Cyg\,X-1. The method output is a mean flare profile showing various substructures. The latter represents stable feature like the central spike and two side-lobes on both sides of the spike. A very similar flare profile was found by \citet{sasada2017} in \kepler\ data of the blazar W2R\,1926+4, and by \citet{dobrotka2019} in the CV MV\,Lyr. Both objects have the central spike and side-lobes with longer time-scales compared to Cyg\,X-1, with blazar time-scales being the longest. All three studies found that characteristic time-scales of the flare substructures are also seen in the corresponding PDS. More interesting it is in the case of MV\,Lyr where both the side-lobes and the central spike has similar characteristic frequencies seen as one single pattern in the PDS. Moreover, use of the averaged flare profile (AFP) as an input to shot noise process simulations, allows to find new PDS structures. Even if such technique is not physically correct, such simulations can be used as a diagnostic tool. \citet{dobrotka2019} used such technique and found a new high frequency component in \xmm\ observations of MV\,Lyr. Recently flare profile analysis was used by \citet{bhargava2022} for study of \astrosat\ and \nicer\ observations of Cyg\,X-1. The author used the AFP for flare-phase-resolved spectroscopy and concluded that the inner edge of the accretion disc moves inwards and outwards as the flares rise and decay. Recently \citet{ninoyu2024} used AFP method to divide the flare into several pre- and past-flare portions to evaluate the polarization evolution during the flare in \ixpe\ data of Cyg\,X-1. The degree of polarization decreased in the segment including and immediately following the flare peak. The authors suggest that the reduced polarization is due to closer proximity of the gas to the BH, therefore the accretion disc contracts with increasing X-ray luminosity of the flare. Similar approach was used also outside of the flickering or accretion system community. \citet{mendoza2022} modelled stellar flares and used similar technique of superposition of individual flares in order to get a mean flare shape. Apparently, the AFP offers additional information.

In this paper we perform flare profile analysis of three selected AGNs and one CV. Due to superior data quality and quantity we focus on IRAS\,13224-3809 in more details. We complete the AFP analysis by autocorrelation function (ACF) study, and we search for similarities between this particular AGN and CV MV\,Lyr.

\section{Studied objects and data}

We selected three Seyfert\,1 type AGNs with count rates high enough, and total duration $T$ long enough for our analysis; IRAS\,13224-3809\footnote{ObsIDs: 0110890101, 0673580101, 0673580201, 0673580301, 0673580401, 0780560101, 0780561301, 0780561401, 0780561501, 0780561601, 0780561701, 0792180101, 0792180201, 0792180301, 0792180401, 0792180501, 0792180601} ($T = 2.1$\,Ms), 1H\,0707-495\footnote{ObsIDs: 0673580101, 0673580201, 0673580301, 0673580401, 0780560101, 0780561301, 0780561401, 0780561501, 0780561601, 0780561701, 0792180101, 0792180201, 0792180301, 0792180401, 0792180501, 0792180601} ($T = 1.4$\,Ms) and Mrk\,766\footnote{ObsIDs: 0096020101, 0109141301, 0304030101, 0304030301, 0304030401, 0304030501, 0304030601, 0304030701, 0763790401} ($T = 0.7$\,Ms).

IRAS\,13224-3809 is a nearby (z=0.066) and X-ray bright galaxy (\citealt{pinto2018}). This object was very intensively observed by \xmm\ yielding the longest data set for an AGN ever. Such an unprecedented high quality observation yields several important results; finding of flux dependence to X-ray time lags (\citealt{fabian2013}), observation of a reverberation signal at lower fluxes (\citealt{kara2013}), discovery of ultrafast outflow (\citealt{pinto2018}) and its flux dependence (\citealt{parker2017}), or finding of soft continuum with strong relativistic reflection and soft excess in the spectrum (\citealt{ponti2010,fabian2013,chiang2015,jiang2018}). This object is one of the most variable AGN in X-rays. Detailed study of this variability performed by \citet{alston2019}. The authors found  multicomponent PDSs with break frequencies around log($f$/Hz) = -4 and -3 with a narrow Lorentzian component also close to log($f$/Hz) = -3. Dependence of the PDS shape on energy and the flux state was discovered too. However, the variability in UV has a low fractional variability amplitude on low frequencies, and no significant correlation with X-rays (\citealt{buisson2018}). Such correlation is expected in high accretion rate Seyferts (see e.g. \citealt{mchardy2016,buisson2017}).

1H\,0707-495 is another AGN with strong fast variability (see e.g. \citealt{pan2016,zhang2018}). The authors reported highly significant quasi-periodic oscillations (QPOs) at about log($f$/Hz) = -3.6 and -3.9. These values satisfy the well known frequency-BH mass relation, which spans from stellar-mass to supermassive BHs. The lack of correlation between X-rays and UV is not unique for IRAS\,13224-3809. \citet{fabian2009} and \citet{pawar2017} showed the same characteristics also for 1H\,0707-495. The latter authors rule out the reprocessing of X-rays as the source of UV variability. Their results are inconsistent with the inward propagating fluctuations in the accretion rate. This AGN shows spectral characteristic similar to IRAS\,13224-3809, i.e. a soft excess and disc reflection (\citealt{fabian2004}). The same was concluded by \citet{ponti2010} who showed that the reflection-based interpretation points towards similarity of IRAS\,13224-3809 and 1H\,0707-495 in terms of both spectral and variability properties.

Observations of Mrk\,766 also showed periodic signals or QPOs detected in X-rays. \citet{page1999} and \citet{boller2001} reported a significant variability with a time-scale of $\sim 5000$\,s and $4200$\,s, respectively. However, \citet{benlloch2001} concluded that this variability of log($f$/Hz) = -3.6 is an artifact of the red noise process. \citet{vaughan2003} found a break frequency at log($f$/Hz) = -3.3 with a broken power law PDS shape similar to the XRB Cyg\,X-1. A similar frequency was also found by \citet{markowitz2007} with strength increasing with energy, a behavior seen in XRBs. A QPO with slightly lower frequency of log($f$/Hz) = -3.8 was discovered by \citet{zhang2017}. Such a QPO can be just transient. The authors pointed toward frequency ratio of 3:2 between this frequency and log($f$/Hz) = -3.6 measured previously. Such a frequency ratio is well-detected in XRBs (see e.g. \citealt{strohmayer2001,remillard2002}). Energy dependent analysis (\citealt{emmanoulopoulos2011}) revealed soft band variations lagging behind hard band ones at frequencies above log($f$/Hz) = -3 making the case similar to 1H\,0707-495.

The data were downloaded from the XMM-Newton Science Archive (XSA), and we obtained light curves with the Science Analysis Software (SAS), version 18.0. Only the soft and hard band light curves of IRAS\,13224-3809 were extracted using newer version 19.0 because they were performed much later for simple chronological reason. We used data from EPIC/pn detector due to its higher throughput. The light curves were extracted from circular region with a radius of 30\arcsec centered on the source and in the energy interval of 0.2 - 10.0\,keV. The backgrounds were extracted from a region offset and the same radius like the source. We used the {\tt epproc} tool to re-generate calibrated event files, and {\tt evselect} for light curve construction. All observation present in \xmm\ archive up to date were used.

MV\,Lyr is a nova-like system, a subclass of CVs (see e.g. \citealt{warner1995} for a review). The accretion discs of these systems in a high state are supposed to be hot with ionised hydrogen, optically thick and geometrically thin. The disc should be developed down to the WD with a very small boundary layer between the inner disc and the WD surface (\citealt{narayan1993}). This is the opposite of the low state with non-ionised cold discs, where the inner disc truncation is expected to be relatively large (see \citealt{lasota2001} for a review). MV\,Lyr is a well studied system with obvious fast variability. First evidence about coherent and QPOs reported \citet{borisov1992}, \citet{skillman1995} and \citet{kraicheva1999}. \citet{scaringi2012a} used extensive \kepler\ data to study the PDS morphology. The authors found four different frequency components with Lorentzian shape in the PDS. \citet{scaringi2013} suggested reprocessing of the X-ray photons on to the accretion disc or inside-out shocks traveling within the disc as the possible source of the fast variability. The model was refined by \citet{scaringi2014}. The author proposed a sandwich model in which the central geometrically thin disc is surrounded by a geometrically thick disc (hot corona). This model was confirmed by \citet{dobrotka2017a} using direct X-ray observation with \xmm.

\begin{figure}
\resizebox{\hsize}{!}{\includegraphics[angle=-90]{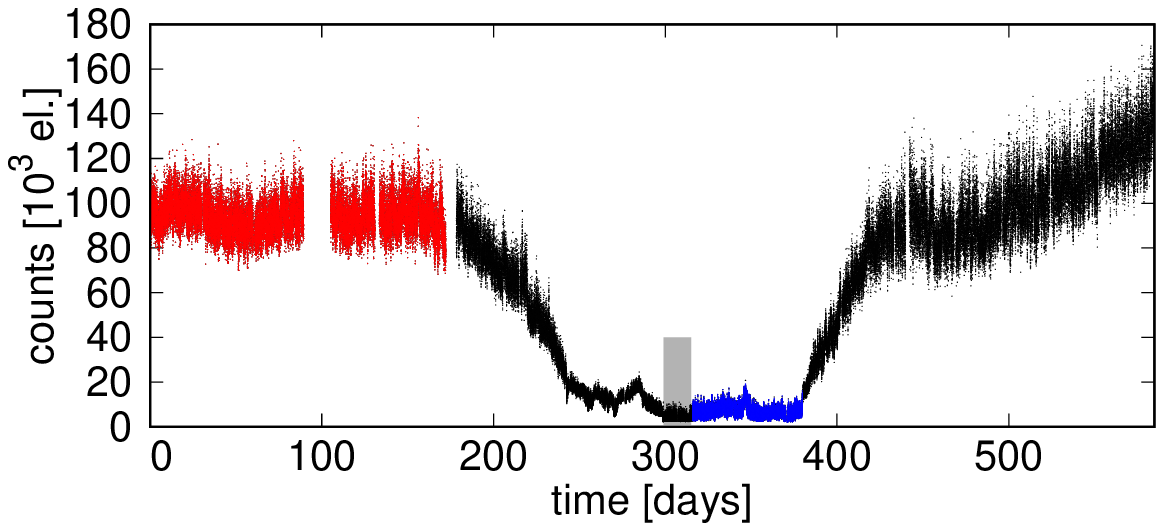}}
\caption{\kepler\ light curve of MV\,Lyr during a transition from high to low state and vice versa. Red points represent the selected high state interval for analysis, while blue points are the selected low state. The shaded area represents the deep low state containing QPOs (see text for details).}
\label{lc_mvlyr}
\end{figure}

For the MV\,Lyr study we used portion of the \kepler\ light curve where the system showed transition from high to low state and vice versa (\citealt{dobrotka2020}). We selected a plateau with constant flux as a high state, and a stable interval during the low state (Fig.\ref{lc_mvlyr}). In the latter a deep low state was identified by \citet{scaringi2017}. In this deep state the systems showed QPOs, and these can contaminate the studied flare profile. Therefore, we excluded this lowest stage from the analysed light curve portion.

\begin{figure}
\resizebox{\hsize}{!}{\includegraphics[angle=-90]{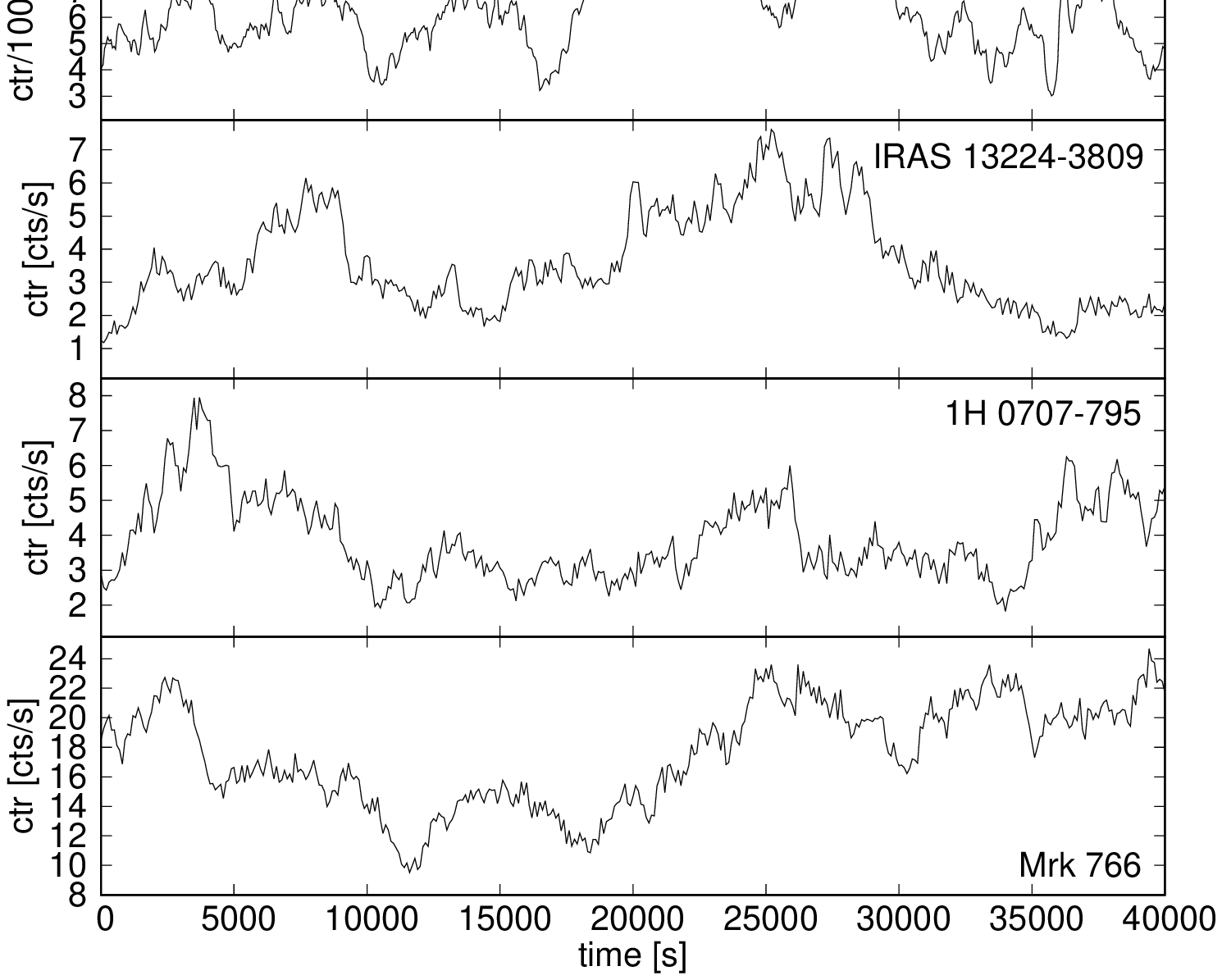}}
\caption{Examples of 0.2 -10.0\,keV light curves for every analysed objects (IRAS\,13224-3809 ObsID 0780561601, 1H\,0707-795 ObsID 0511580101, Mrk\,766 ObsID 0304030501). All light curves are shown from the beginning of the observation except MV\,Lyr in low state. The latter starts from day 350 (see Fig.~\ref{lc_mvlyr}).}
\label{lc_examples}
\end{figure}

Examples of light curves for individual objects are shown in Fig.~\ref{lc_examples}. At first look all light curves have similar characteristics except MV\,Lyr in the high state where well resolved flares with time scale up to about 1\,ks are present.

\section{AFP analysis}

The PDS analysis is a very popular but generalized technique. The inconvenience is that various structures of a time series with the same or a similar characteristic frequency can blend together and form a single PDS feature. Moreover, if the studied time series is time-reversed, the PDS is unchanged. The same is valid for ACF making exact timing location of an event difficult, i.e. if a side-lobe is localized in a rising branch of a flare, the lobe is detected by the ACF but the location information is lost. On the other hand, the AFP method extracts and conserves not only the Fourier amplitude information of flare events from time series data, just like PDS and the ACF, but also the relative phase information in between different frequencies related with the flares. Therefore, we apply the AFP method which yields a result which may be asymmetric and allows to localize relative positions of different sub-structures in time.

The AFP method has also specific disadvantages. In superposing events aligning the peaks, especially for low statistical data, count fluctuations smear fine structures of the profile. Furthermore, higher and (much) lower frequency components unrelated with the selected peak events are completely ignored (e.g., high frequency components at > 1 Hz in Cygnus\,X-1, see \citealt{negoro2001}). Furthermore, if the averaged flares have various types of profiles which appears in the case of non-stationarity, the current AFP does not provide a true average profile, and the average profile must depend on the criteria to select flares. The stationarity is discussed in more details in Section~\ref{section_stationarity}. Finally, use of too short time bins of the light curve can generate a fake central spike. We investigate this strong artifact in the next section.

\subsection{Method}

The goal of the AFP method is "direct" visualisation of individual PDS components in the flare shape, like in the case of Cyg\,X-1 (\citealt{negoro2001}). The whole idea is supported by simulations of radiation generated by an inhomogeneity propagating thru advection dominated flow\footnote{\citet{manmoto1996} did not consider the origin or seed of the inhomogeneity. They only showed what happens if a fluctuation is present in the advection flow. The inhomogeneity can be kind of accretion fluctuation or turbulence.} (\citealt{manmoto1996}). Such inhomogeneity do not dissipate and generates a typical flare profile. Many subsequent inhomogeneities generate many flares which can overlap and are affected by noise. A single flare is not enough for analysis of the profile, because usually all faint features are hidden in the noise. The summation and averaging of individual flares smooth out these contaminations and only the original flare remains\footnote{It is important not to confuse the method with shot noise process where individual flares are summed to get a simulated light curve. Such process is additive which is not consistent with observations. The summation and averaging of many flares in our method is done for smoothing, not for light curve construction.}.

The method has three steps. First is the identification of flares. A light curve data point is identified as a peak if $N_{\rm pts}$ points to the left and $N_{\rm pts}$ points to the right have lower fluxes than the tested point.

In the second step, the flare extension to be averaged must be defined, in this case, as $N_{\rm ptsext}$ points to the left and right from the peak. \citet{dobrotka2019} used $N_{\rm ptsext} = N_{\rm pts}/2$ in order not to superimpose a declining branch of one flare onto a rising branch of the adjacent flare, and vice versa.

The amount of \xmm\ data is much lower than that of the \kepler\ light curve data. Therefore the parameters $N_{\rm pts}$ and $N_{\rm ptsext}$ must be setup with caution. The condition $N_{\rm ptsext} = N_{\rm pts}/2$ shortens the time extension of the studied flare. This must be compensated by larger values of $N_{\rm pts}$. However, larger $N_{\rm pts}$ decreases the number of flares per AFP, which increases the noise. \citet{dobrotka2019} showed that the condition $N_{\rm ptsext} = N_{\rm pts}$ is still acceptable, and any method artifacts occur for $N_{\rm ptsext} > N_{\rm pts}$. Therefore, for this work we use $N_{\rm ptsext} = N_{\rm pts}$. This empirical test explains also why overlapping flares do not contaminate the shape of the AFP. Even if individual selected flares are certainly contaminated by other overlapping flares (see Fig.~3 in \citealt{dobrotka2019}), those are redistributed randomly and summation of many flares smooth out any randomly redistributed events. Only regularly repeating feature will be seen in the AFP.

In the last step after the flare selection, the flare data points are averaged with maxima aligned, and the resulting averaged flux minimum value is subtracted from all averaged points. All flares with rare individual null points or missing data (at the edge of the light curve or gaps) were excluded from the averaging process.

The method is described in more details with several tests like role of $N$ parameter on results, sampling of \kepler\ data with instrumental interruption, etc. in \citet{dobrotka2019}. The authors performed also several reality tests like comparison of the AFP with ACF, and found the same features in both methods.

For \xmm\ EPIC/pn data of the selected AGNs we used 100\,s bin as an empirical compromise between S/N ratio and resolution. As parameter $N_{\rm pts}$ we used 20 and 40 yielding time extension of 2000 and 4000\,s on both sides from the center of the averaged flare.

For \kepler\ data of MV\,Lyr we used $N_{\rm pts} = 34$ and 68 to get similar time extension like for the AGNs. With the \kepler\ cadence of 58.8\,s the corresponding time extension are 1999.2 and 3998.4\,s. As discussed in \citet{dobrotka2019} the long-term trend and barycentric correction of \kepler\ data do not affect the result.

\begin{figure}
\resizebox{\hsize}{!}{\includegraphics[angle=-90]{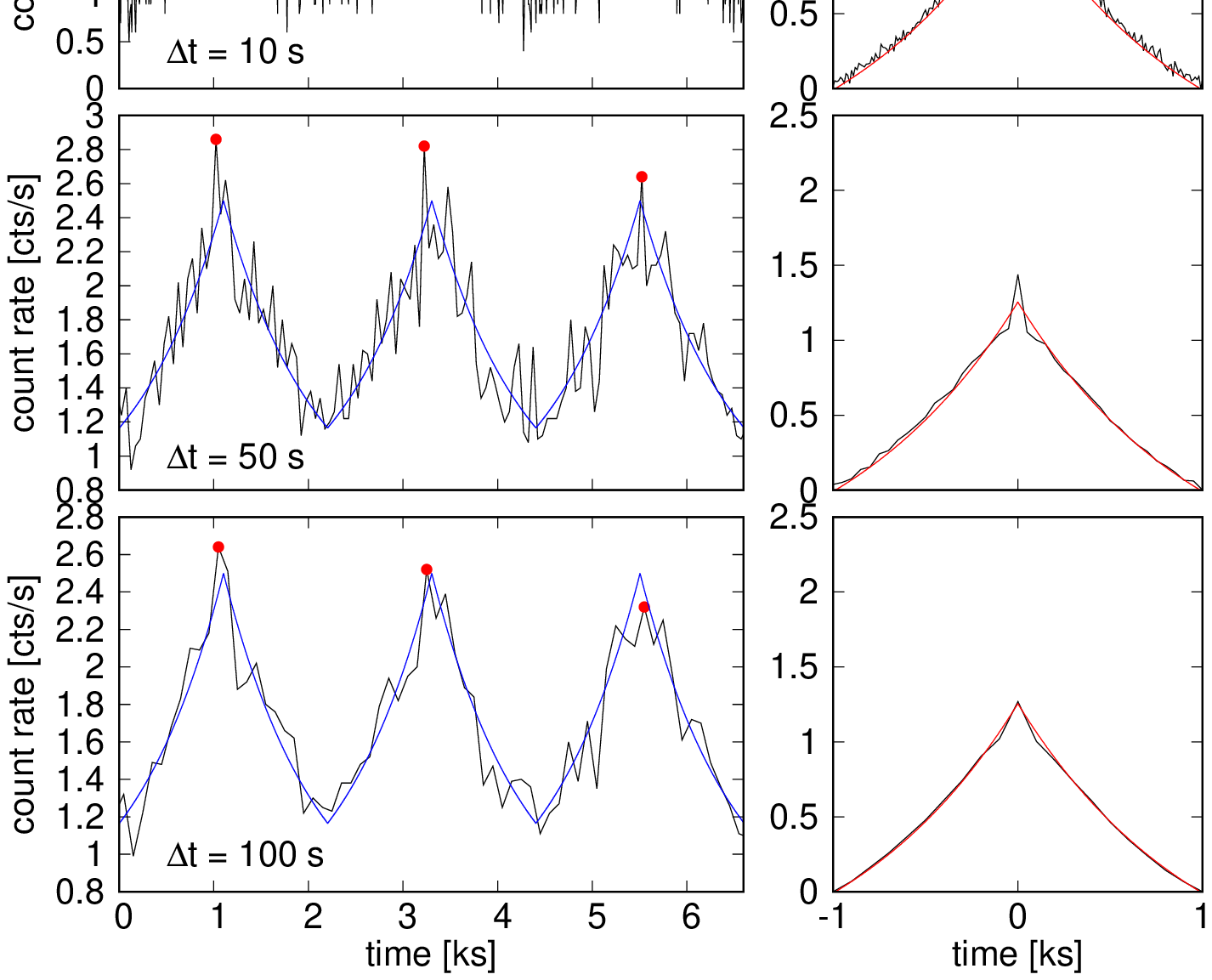}}
\caption{Simulated light curves with AFPs. Left panels - light curves with synthetic flares using Equation~(\ref{correct_fit}) and $F_{\rm c}$ = 2\,cts/s, $F_{\rm c}$ = 0.5\,cts/s, $T_{\rm r} = T_{\rm d} = 1000$\,s with Poisson noise. Different time bins $\Delta t$ are shown. The detected flare "maxima" are shown as red points. Blue is the original light curve without noise. Right panels - AFPs calculated from synthetic light curves with 200 flares. Red line depicts the reference light curve without noise.}
\label{profile_spike_bias}
\end{figure}

Finally, important is to exclude one selection effect affecting the resulting profiles. Since the method first searches for a maximum, this maximal point of a flare can be a scattered point due to count fluctuations toward a higher count rate. Such a selection effect results in a deviated central point. In Fig.~\ref{profile_spike_bias} we show simulations of flares using Equations~\ref{correct_fit}. We used different time bins to show the role of the Poisson noise. Apparently, the method detects the most deviated point and not always the true flare maximum. Such detection of most deviated point results in accumulation of this systematic bias and a central fake spike appears. Larger is the time bin, smaller is the spike bias. Since the method detects the most deviated point because due to Poisson noise, this bias is inherently present in any observed light curve. Moreover, since the method is not detecting the real flare maximum, the detected maxima are scattered around the real value. It is easy to demonstrate by simulations, that larger is the time bin, lower is this scatter. The simulations in Fig.~\ref{profile_spike_bias} show that 100\,s time bin yields results where the central spike bias is negligible. But this is valid only for the performed simulations with specific parameters. In reality it depends on flare amplitude and mean count rate, i.e. parameters affecting the Poisson noise level. Since this artificial count increase is affecting only a single central point, any real substructure like real central spike must be represented by more than one single point. The single central point consists of real flux level of the flare maximum and the Poisson noise artifact. With the recent method it is impossible to separate the real flux level and the fake contribution.

Moreover, in the case of shortest time bins, even two maxima can be detected for one single flare if the count rate is the same (top panel of Fig.~\ref{profile_spike_bias}). Another artifact is some flattening of the averaged flare close to the central fake spike. This effect is negligible in simulations with time bin of 100\,s. Apparently, this flattening is related to the central spike bias and is probably generated by non aligned flares before averaging. Time bin large enough reduces the central spike bias and also the central flattening.

Therefore, we used 100\,s bin for all X-ray light curves as a compromise between scatter and time resolution. In order to exclude any potential central spike bias when comparing individual AFPs, we excluded the central point from all profiles since the scatter and flare amplitudes are different from source to source. Subsequently we normalize the averaged flare flux by its remaining maximal value. All other analysis like comparison of individual amplitudes in a single source or profile fitting we performed in standard count rates without any modification, just the central point is not taken into account during the fitting procedure.

On the other hand, the exclusion of the central point can have negative impact. It smears possible short-term variations associated with the flare. A solution would be to identify flares in a different way. For example finding the highest point can be just a step to identify possible flare maximum, and the true flare can be selected by fitting of a typical profile (described in the next section). However, as mentioned and demonstrated in Fig.~3 of \citet{dobrotka2019} the flares does not have typical profile because of contamination by adjacent flares. The shapes are very different with sharper and wider peaks, clear decreasing/increasing branches or variable branches with strong humps or plateau with constant count rate, etc.. Our situation is very different from the case of fast variability in the BH in the hard state (\citealt{negoro1994,focke2005}) and blazar (\citealt{sasada2017}), or stellar flares (\citealt{mendoza2022}) and gamma-ray bursts (\citealt{sakamoto2007}) where the flares are relatively isolated. Those are likely suitable for fitting with a typical profile. The difference in flare behaviour is not only a technical challenge or difficulty. It says something about the origin of the variability and physical conditions in the accretion flow.

The central region importance is well seen in stellar flare shape where it has been shown that the flare profile does not have break points between the rise and decay, but instead a continuous model where peaks roll over are needed (\citealt{kowalski2016,jackman2018,jackman2019,howard2022}). Nevertheless, in this work, we only focus on wider patterns like side-lobes or central spike, rather than the peak itself. Therefore, we accept the exclusion of the central point because it does not affect the studied flare features.

\subsection{Results}
\label{section_results}

\begin{figure}
\resizebox{\hsize}{!}{\includegraphics[angle=-90]{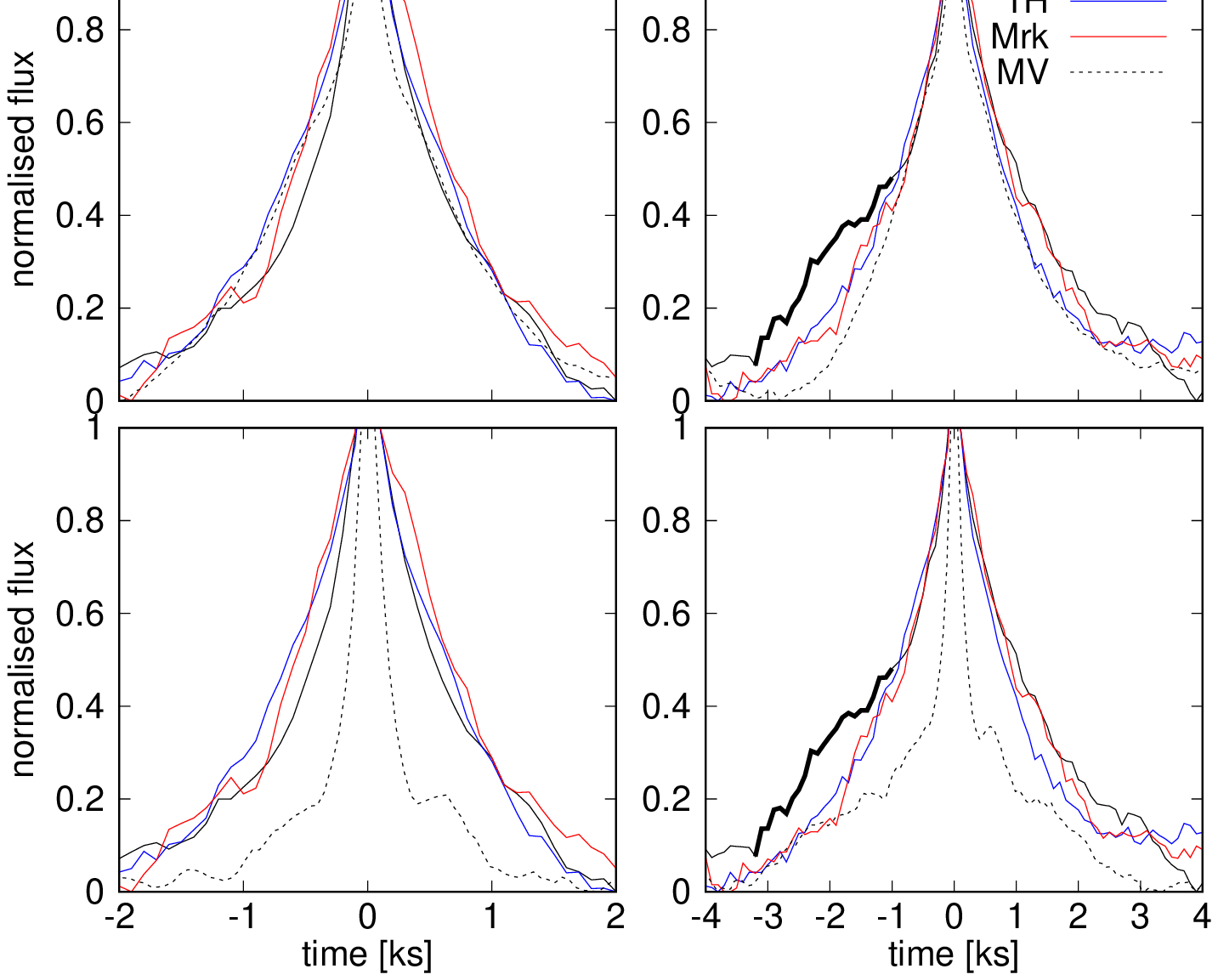}}
\caption{AFPs for two different time extensions (from -2 to 2\,ks in the left panels, from -4 to 4\,ks in the right panels). The three AGNs are compared to CV MV\,Lyr. The latter is shown in two different states; in a low (upper panels) and high (lower panels) state. The averaged number of flares is summarized in Table~\ref{flares_stat}. Black thick line shows the side-lobe between -3.2 and -1\,ks of IRAS\,13224-3809 discussed in the text.}
\label{profiles}
\end{figure}

The resultant AFPs are depicted in Fig.~\ref{profiles}, and the numbers of averaged flares are summarized in Table~\ref{flares_stat}. Clearly all the AGN profiles are very similar except the IRAS\,13224-3809 case showing noticeable side-lobe between -3200 and -1000 seconds. The most important result is the indistinguishable resemblance of the MV\,Lyr case in the low state, while the profile is totally different in the high state. Apparently the central spike in the CV and the side lobe(s) make the case different\footnote{We note that the central spike of MV\,Lyr consists of more than one single central point, therefore it is a real structure.}.
\begin{table}
\caption{Summarization of number of averaged flares for individual time extensions.}
\begin{center}
\begin{tabular}{lcrr}
\hline
\hline
object & state & 4000\,s & 8000\,s\\
\hline
MV\,Lyr & high & 2631 & 1377\\
& low & 978 & 574\\
IRAS\,13224-3809 & & 327 & 158\\
1H0707-495 & & 184 & 103\\
Mrk\,766 & & 92 & 47\\
\hline
\end{tabular}
\end{center}
\label{flares_stat}
\end{table}

\begin{figure}
\resizebox{\hsize}{!}{\includegraphics[angle=-90]{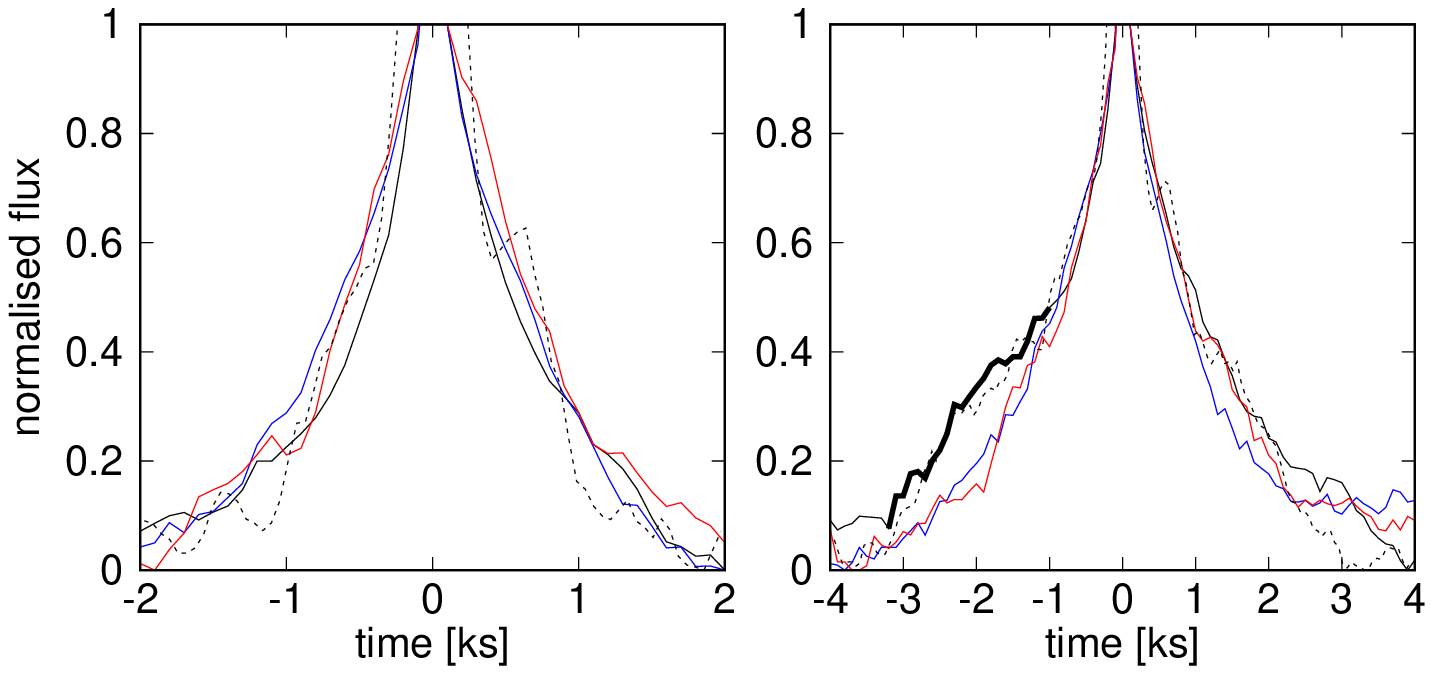}}
\caption{The same as bottom panels of Fig.~\ref{profiles} with MV\,Lyr case multiplied by 3 (left panel) and 2 (right panel).}
\label{profiles_mvmodif}
\end{figure}

Multiplying the MV\,Lyr profiles in the high state by a factor of a few lifts them up. As a result the central spike is out of the figure and we can compare only the wider base of the CV with those of the AGNs. The left panel of Fig.~\ref{profiles_mvmodif} shows that the profile more or less agrees with the AGN profiles. However, more interesting is the right panel showing larger time extension. Not only the overall profile matches the AGN cases, but also the left side-lobe matches similar structure seen in IRAS\,13224-3809. Since the MV\,Lyr profile is calculated from 2631 flares, this structure is unambiguous.

\begin{figure}
\resizebox{\hsize}{!}{\includegraphics[angle=-90]{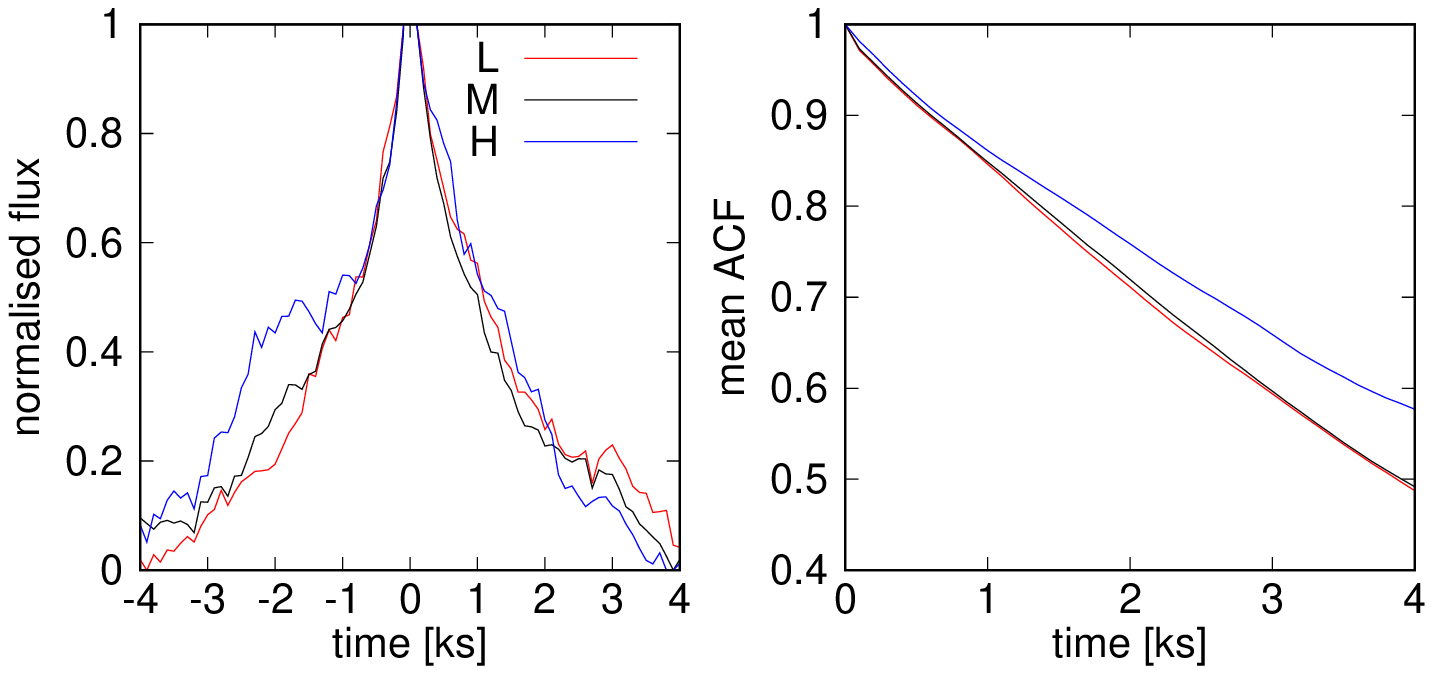}}
\caption{AFPs of IRAS\,13224-3809 with corresponding ACFs. Left panel: the same as the upper right panel of Fig.~\ref{profiles} but the IRAS\,13224-3809 flares are calculated from low ("L"), low-medium ("M") and high ("H") flux subsamples of the IRAS\,13224-3809 light curve. Right panel: mean ACFs of the same light curve subsamples of IRAS\,13224-3809.}
\label{profiles_lf_hf}
\end{figure}

\citet{alston2019} and \citet{caballero2020} divided the IRAS\,13224-3809 light curve into low, medium, and high states. We did the same and for simplicity we divided the light curve into high, low-medium and low flux data. Three observations have remarkably higher mean flux (higher than 4\,cts/s) and amplitude of variability (square root of the variance). These are classified as high flux data. The observations with mean flux lower than 2.5\,cts/s represents low flux data, and the rest are low-medium data. Left panel of Fig.~\ref{profiles_lf_hf} compares the averaged flares from all three light curve subsamples. We focus on the largest time extension from -4 to 4\,ks where IRAS\,13224-3809 shows a side-lobe on the rising part of the flare. Clearly, this side-lobe is dominant in the high flux flare. The lower is the flux, the closer is the averaged profile to other three objects. In general the high flux case tends to be wider in time interval from -3 to 2\,ks, while the low flux case matches well with the other objects.

\begin{figure}
\resizebox{\hsize}{!}{\includegraphics[angle=-90]{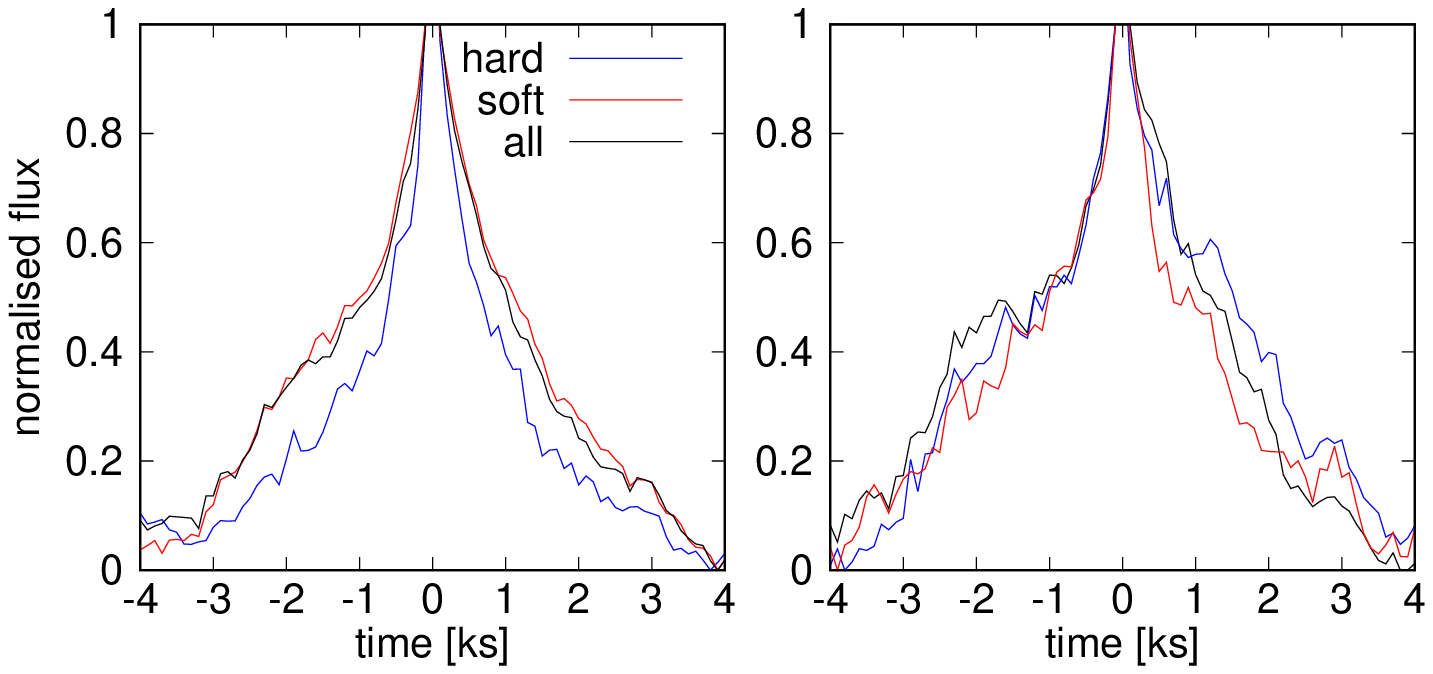}}
\caption{Comparison of IRAS\,13224-3809 AFPs in soft and hard bands together with that from the whole energetic interval. Left panel represents the same as Fig.~\ref{profiles}, while right panel compares only high flux interval like in Fig.~\ref{profiles_lf_hf}.}
\label{profiles_soft_hard}
\end{figure}

In order to study the energetics of the flares we divided the light curve into soft (below 1\,keV) and hard (above 1\,keV) bands. Majority of photons is below 1\,keV. Therefore, if we would like to have the same number of photons in both bands, we would need to use less than 0.6\,keV as a threshold energy. This is too low for a meaningful energetic study\footnote{We tried such two bands, and we did not see significant difference between soft and hard AFPs.}. The comparison of IRAS\,13224-3809 AFPs in the different energy bands are in Fig.~\ref{profiles_soft_hard}. We focus again on the largest time extension where IRAS\,13224-3809 shows a side-lobe on the rising part of the flare. When using the whole light curve the flare width in the hard band is narrower than that in the soft band, and the side-lobe on the rising branch is not seen (left panel of Fig.~\ref{profiles_soft_hard}). When comparing only high flux flares from Fig.~\ref{profiles_lf_hf} the corresponding hard band is slightly wider, but there is no significant difference in the side-lobe (right panel of Fig.~\ref{profiles_soft_hard}). This means that the side-lobe is present in both bands during the high flux period, and absent in the lower flux periods.

The meaning of the AFP is the possibility to see the time-scales "directly". The individual profiles are best described by the following exponential function (\citealt{sasada2017,dobrotka2019})
\begin{equation}
F(t) = \left\{
\begin{array}{ll}
F_{\rm c} + F_{\rm 0} e^{t/T_{\rm r}}, & t < 0\\
F_{\rm c} + F_{\rm 0} e^{-t/T_{\rm d}}, & t > 0,
\label{correct_fit}
\end{array}
\right.
\end{equation}
where $t$ is time, $T_{\rm r}$ and $T_{\rm d}$ are time constants of a rising and declining branch, respectively. These time constants correspond to knees in the PDS with frequencies of $1/(2 \pi T_{\rm r})$ and $1/(2 \pi T_{\rm d})$. $F_{\rm c}$ and $F_{\rm 0}$ represent the constant level and the amplitude of the flares, respectively. We excluded the central point at 0\,s from the fitting for the reasons described above.

\begin{figure}
\resizebox{\hsize}{!}{\includegraphics[angle=-90]{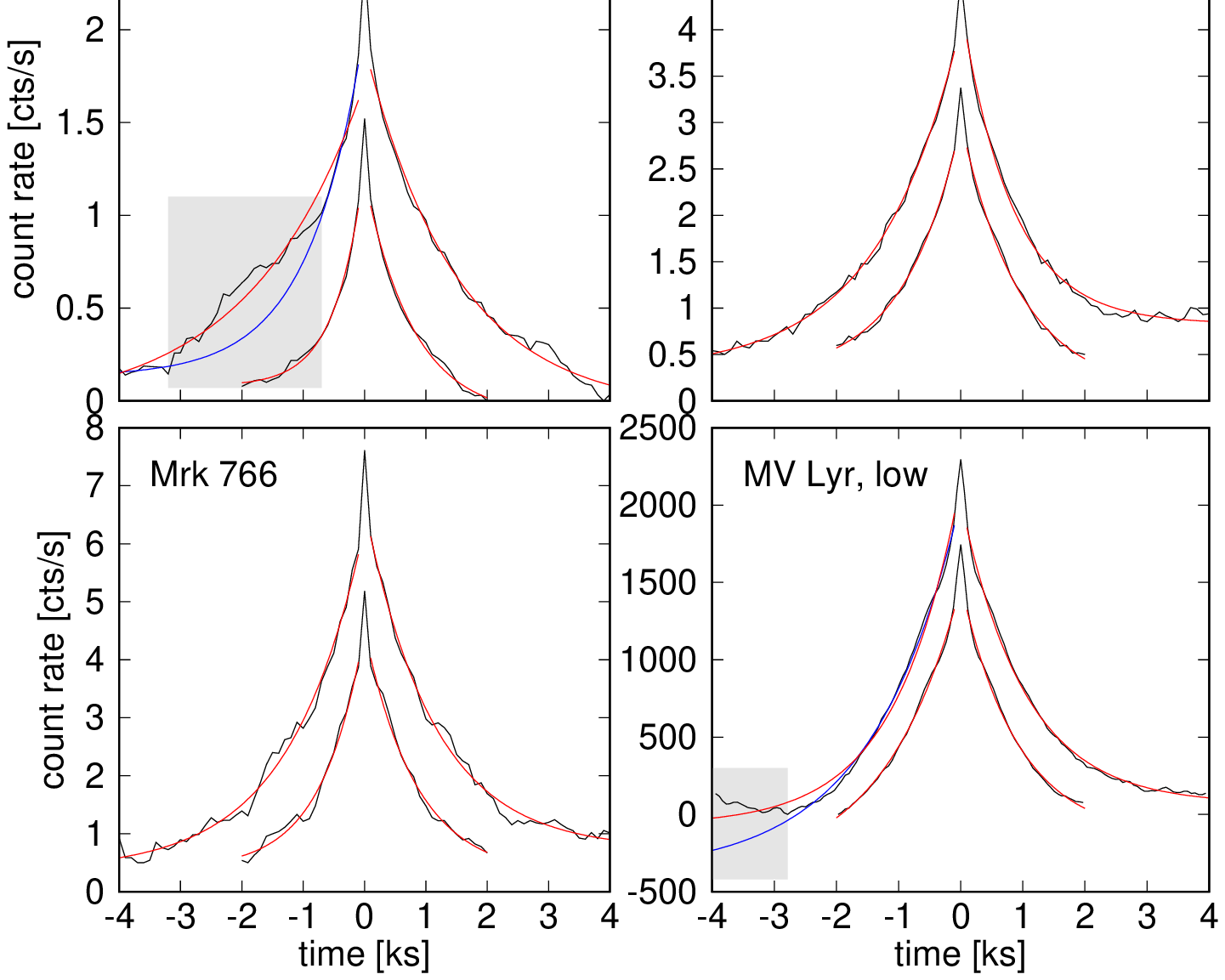}}
\caption{Fits to the averaged profiles from Fig.~\ref{profiles} using Equation~(\ref{correct_fit}). Red lines represent the fits to the whole time extension, while blue lines are fits after some time interval exclusion (shaded areas).}
\label{profile_fits}
\end{figure}

\begin{figure}
\resizebox{\hsize}{!}{\includegraphics[angle=-90]{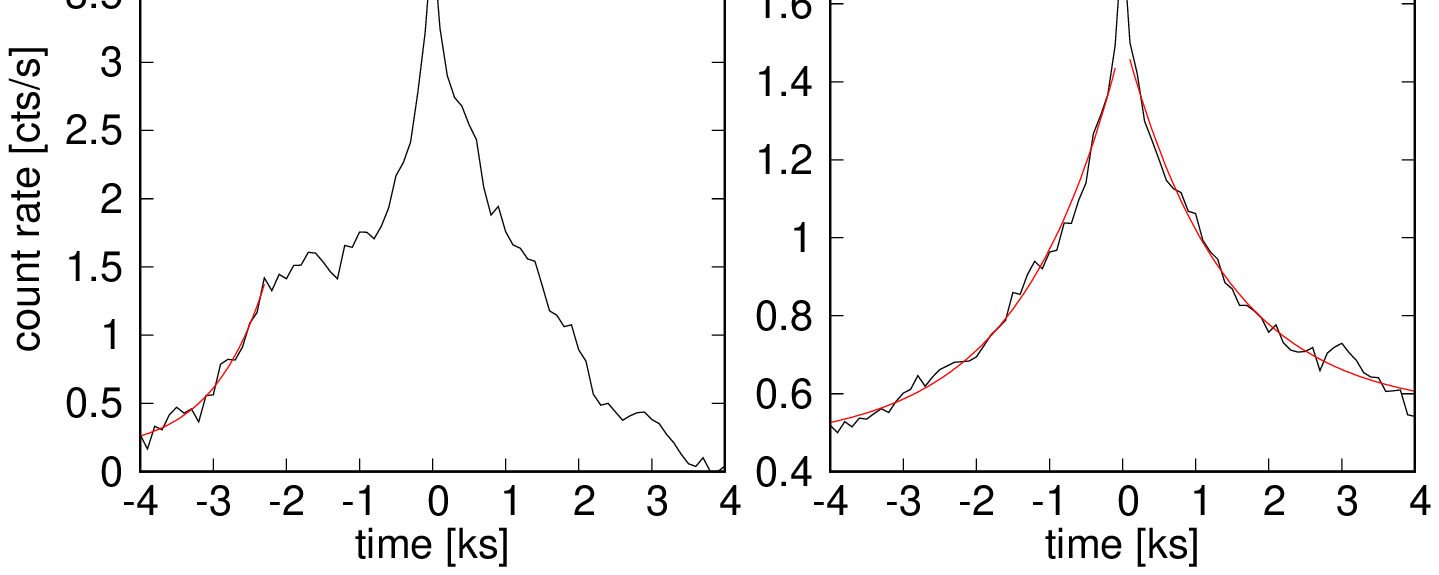}}
\caption{The same as Fig.~\ref{profile_fits} but only for high (left panel) and low (right panel) flux flares of IRAS\,13224-3809.}
\label{profile_fits_iras}
\end{figure}

The fits are shown in Fig.~\ref{profile_fits}, and the parameters are summarized in Table~\ref{fit_param}. The first fits of the flares with time extent of $2 \times 2000$\,s ($N_{\rm pts}$ = 20) describes all the profiles well. On the other hand, there is a problematic structure in the time extent of $2 \times 4000$\,s ($N_{\rm pts}$ = 40). Strong deviations from the model function appear in the case of IRAS\,13224-3809. The rising branch does not have a smooth development, and it has the side-lobe mentioned above. We selected a region where this lobe is clearly seen and we excluded this time interval from the fitting procedure (shaded area in Fig.~\ref{profile_fits} and blue line). Apparently the two fits yield different time scales in the case of IRAS\,13224-3809. Therefore, it is better to treat this case in more details. Fig.~\ref{profile_fits_iras} shows the high and low flux version of the flares with fits. In the former we fitted only the rising part of the flare on selected time interval. The fit describes well the beginning of the side-lobe. Finally, the low flux AFP is well fitted without any significant deviation.

Problematic fitting behaviour is also seen in the rising branch of the MV\,Lyr AFP in the low state. The profile rises from approximately -3000\,s toward lower time values. We selected this time region and excluded it from the fitting procedure (shaded area in Fig.~\ref{profile_fits} and blue line).
\begin{table*}
\caption{$T_{\rm r}$ and $T_{\rm d}$ parameters and corresponding PDS characteristic frequencies $f_{\rm r}$ and $f_{\rm d}$ for individual objects. $t_{\rm ext}$ is the time extension of the averaged flare.}
\begin{center}
\begin{tabular}{lccccc}
\hline
\hline
object & $t_{\rm ext}$ & $T_{\rm r}$ & $T_{\rm d}$ & log($f_{\rm r}$/Hz) &log($f_{\rm d}$/Hz)\\
& (ks) & (s) & (s) & &\\
\hline
IRAS\,13224-3809 & 4 & $473.6 \pm 21.6$ & $794.5 \pm 52.2$ & -3.47 & -3.70\\
1H\,0707-495 & 4 & $911.8 \pm 44.8$ & $950.2 \pm 55.2$ & -3.76 & -3.78\\
Mrk\,766 & 4 & $691.5 \pm 66.1$ & $844.4 \pm 64.4$ & -3.64 & -3.72\\
MV\,Lyr & 4 & $1232.2 \pm 54.8$ & $925.8 \pm 37.1$ & -3.89 & -3.76\\
\hline
IRAS\,13224-3809 & 8 & $1838.1 \pm 212.4^a$ & $1495.0 \pm 74.0$ & -4.06 & -3.97\\
& & $898.2 \pm 43.1^b$ & & -3.75 &\\
& & $1368.8 \pm 69.4^c$ & $1361.6 \pm 86.23^c$ & -3.93 & -3.93\\
& & $731.7 \pm 129.2^d$ & & -3.66 &\\
1H\,0707-495 & 8 & $1330.3 \pm 31.7$ & $826.0 \pm 25.7$ & -3.92 & -3.72\\
Mrk\,766 & 8 & $1208.9 \pm 58.4$ & $1123.9 \pm 43.4$ & -3.88 & -3.85\\
MV\,Lyr & 8 & $1026.3 \pm 42.0^a$ & $1027.4 \pm 16.4$ & -3.81 & -3.81\\
& & $1418.6 \pm 48.4^b$ & & -3.95&\\
\hline
\end{tabular}
\end{center}
$^a$ fits on the whole time extension and without any alternation\\
$^b$ fits on altered time interval\\
$^c$ fits of low flux version of the flare\\
$^d$ fits of side-lobe of high flux version of the flare
\label{fit_param}
\end{table*}
%
%
%

Apparently none of the characteristic frequencies reaches log($f$/Hz) = -3 seen in the high state of MV\,Lyr or in IRAS\,13224-3809 (\citealt{alston2019}). To get such a frequency from the exponential rise or decay profile, the $T_{\rm r}$ or $T_{\rm d}$ parameter should be about 160\,s.  The time scale of 160\,s corresponds to only two time bins, and it requires to use $N_{\rm pts}$ = 6 or less for averaging the IRAS\,13224-3809 flares, and fitting up to or from 0\,s (central point included). Therefore, the very central part of the spike has too low resolution for detection of the log($f$/Hz) = -3 frequency (if present).


The fitted $T_{\rm r}$ and $T_{\rm d}$ parameters varies from object to object, and the time constants depend on the time extension to be investigated. All the corresponding frequencies log($f$/Hz) are, however, in the range of $-3.5$ to $-4.1$ including MV\,Lyr. We also note that the average time constants of $T_{\rm r}$ and $T_{\rm d}$ in 1H\,0707-495 and MV\,Lyr are different only less than 20\% in the two time extension fits, implying these values are characteristic time constants of these objects. Indeed, $T_{\rm r} + T_{\rm d}$ ($\sim 2000$\,s) of 1H\,0707-495 is about half of 3800\,s QPO periodicity observed in the \xmm\, observations (\citealt{pan2016,zhang2018}).

\begin{figure}
\resizebox{\hsize}{!}{\includegraphics[angle=-90]{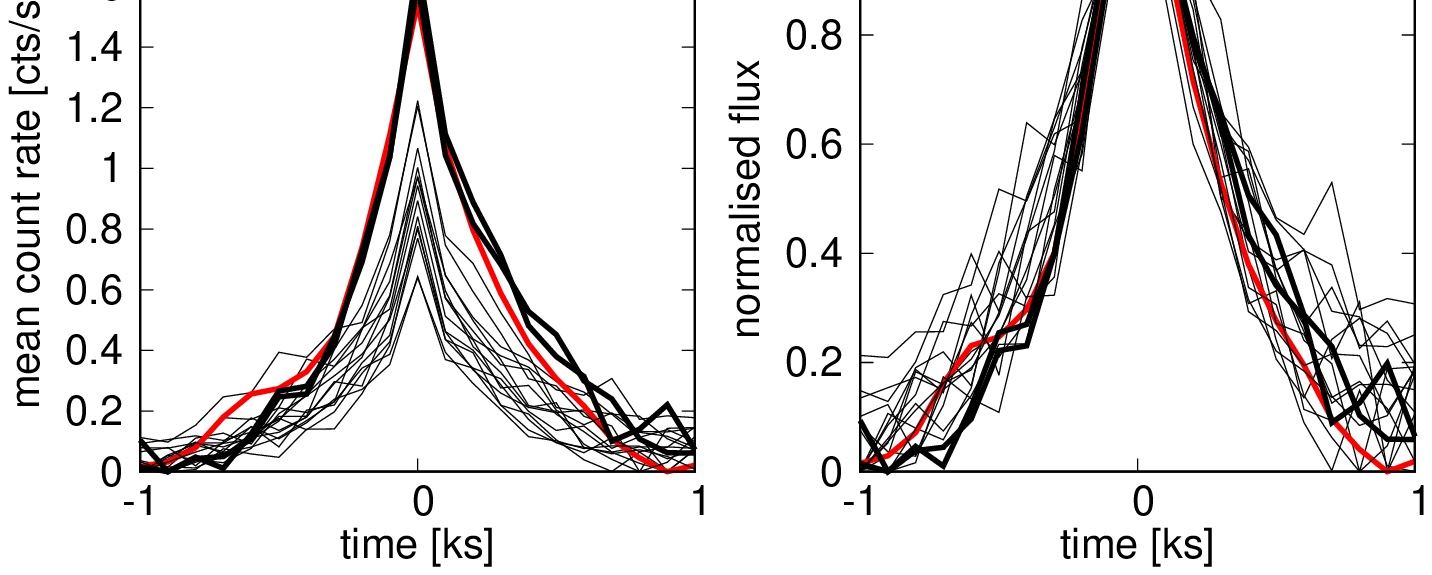}}
\caption{AFPs from individual observations. The thick lines represent the ObsIDs: 0780561601 (red), 0792180401 (black) and 0792180601 (black). Left panel represents the true amplitudes. Right panel shows the same comparison like in Fig.~\ref{profiles} where the central point is excluded and the resulting maxima are set to 1.}
\label{profiles_individual}
\end{figure}

Finally, it is worth to compare AFPs from individual observations. To increase the number of averaged flares we used time extension of only 2000\,s (from -1000 to 1000\,s, $N_{\rm pts}$ = 10). Individual profiles are shown in Fig.~\ref{profiles_individual}.

All profiles have very similar amplitudes except three cases which are considerably higher (shown as thick lines). All three observations are the ones already mentioned with the highest mean fluxes and amplitudes of variability. The three averaged profiles are very similar. Some minor deviations of the ObsID: 0780561601 case shown as red line are possible at the wings below -500 and above 200\,s. The two other profiles (ObsIDs: 0792180401 and 0792180601) represented by black lines are almost identical. However, due to very low number of averaged flares (48 red, 45 and 49 black) we can not say whether such faint details are real or not. Moreover, as shown by lower panel of Fig.~\ref{profiles_individual} all profiles are comparable, just the scatter of low amplitude cases is too large due to low number of counts. The conclusion is that at least the three dominant profiles show similar slopes or curvatures without any obvious substructure like side-lobe in the time interval of -1000 to 1000\,s.

\subsection{Comparison with ACF}

\citet{dobrotka2019} compared the AFP with ACF, and found the same features in both methods. We performed the same confidence test.

As ACF requires continuous uninterrupted light curves, we calculated ACFs of individual \xmm\ observations and averaged them. MV\,Lyr data were divided into subsamples separated by larger gaps. This resulted in three and six subsamples for low and high state, respectively. Sporadic zero points were removed and replaced by median values using Hampel filter\footnote{Python hampel library \url{https://github.com/MichaelisTrofficus/hampel_filter}.}

\begin{figure}
\resizebox{\hsize}{!}{\includegraphics[angle=-90]{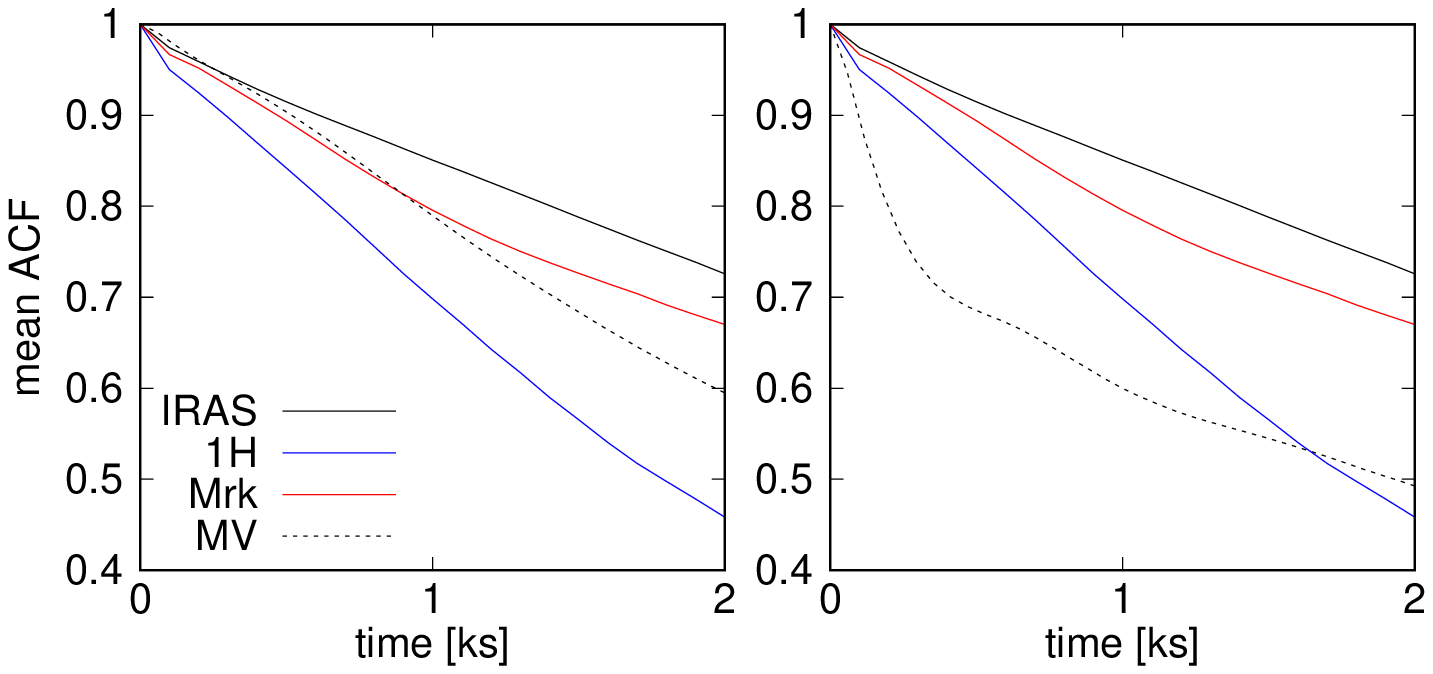}}
\caption{Comparison of mean ACFs. Left panel compares AGN ACFs with ACF of CV MV\,Lyr in a low state, while the right panel shows MV\,Lyr in a high state.}
\label{acf_comparison}
\end{figure}

The differences in ACFs are clear in Fig.~\ref{acf_comparison}. The mean ACFs of AGNs and CV in the low state are smooth and monotonic. Contrary, the CV in high state shows structured ACF with a steep decline followed by small hump representing the central spike and side-lobes, respectively. After the small hump the ACF continues monotonically as in the AGN cases.

The same confidence test can be applied to the side-lobe in the rising part of the flare depicted in right panel of Fig.~\ref{profiles_lf_hf}. Mean ACFs calculated from the same light curve subsamples are depicted in the left panel of Fig.~\ref{profiles_lf_hf}. The difference between low and low-medium flux AFPs is small, but the difference is already seen. The same is valid for corresponding ACFs. The biggest difference is seen in high flux profile which agrees with highest difference in ACFs. The studied side-lobe starts to be deviated from other profiles at approximately -1\,ks. The ACFs are also deviated from approximately 1\,ks. Apparently, the ACFs support the existence and flux dependance of the side-lobe detected in the AFP of IRAS\,13224-3809.

\section{Discussion}

We performed AFP analysis of three selected AGNs observed with \xmm. We compare the results with AFP of the CV MV\,Lyr observed with \kepler\ spacecraft.

One of the selected AGNs is IRAS\,13224-3809. This AGN has the longest observation list in the \xmm\ archive which makes it suitable for more detailed analysis. We searched for characteristic frequencies and for any relation between the AFP and the standard PDS.

\subsection{Stationarity}
\label{section_stationarity}

Our AFP method selects flares and averages them in order to get a mean typical profile from the whole light curve. Therefore, the method is ideal for stationary processes in which the shape of the flare does not change in time. \citet{alston2019} reported that the \xmm\ data of IRAS\,13224-3809 show evidence for non-stationarity. The authors compared PDSs calculated from several epochs. They concluded that the PDSs are well described by two peaked components, and the non-stationarity is seen as clear changes in PDS normalisation as well as smaller changes in PDS shape, i.e. shift in peak frequency of the low frequency component close to log($f$/Hz) = -4.3. If just the normalisation (power) of PDS features changes, any break or QPO frequency should be unchanged, and this is what we search for in the AFP.

However, the smaller shift in the low frequency component can affect the AFP. But the changes are only small, and \citet{alston2019} concluded that the PDSs have similar shapes from epoch to epoch, and the variability components can't be varying dramatically. Therefore, we investigated individual observations of IRAS\,13224-3809, and obtained similar AFPs showing some scatter due to not enough of counts except three cases (Figs~\ref{profiles_individual}). These three cases from different epochs show very comparable AFPs. This implies, that either all three observations with higher flux have comparable characteristics, or any non-stationarity within different epochs has no significant impact on the AFP. Any change in PDS components is hidden in the noise of the AFPs.

\subsection{Comparison of AFP substructure with the PDS in IRAS\,13224-3809}
\label{discussion_comparison}

The remarkable observations of IRAS\,13224-3809 by \xmm\ allow us to study the AFP in more details and its evolution in time. First, we want to point out similarities between our AFP analysis and PDS studied by \citet{alston2019}. Our AFP fitting revealed potential characteristic frequencies at log($f$/Hz) = -3.47, -3.70, -3.97 or -4.06 (Table~\ref{fit_param}). These values correspond to bend frequency interval from $8.7 \times 10^{-5}$ to $3.4 \times 10^{-4}$\,Hz in the 1.2-3\,keV and 3-10\,keV bands seen as blue and gray histograms in Fig.~10 (panel a) of \citet{alston2019} using bend$_1$ + bend$_2$ + lor$_1$ model. The same energy band of 1.3-3\,keV shows also a bend frequency at approximately $1 - 2 \times 10^{-3}$\,Hz seen as blue histogram in Fig.~10 (panel b) of \citet{alston2019}. This bend frequency suggests a time constant of $\sim 100$\,s in Equation~(\ref{correct_fit}). For such high frequencies we apparently lack resolution. These imply that our AFP involves the low frequency properties of the observed PSD.

\begin{figure}
\resizebox{\hsize}{!}{\includegraphics[angle=0]{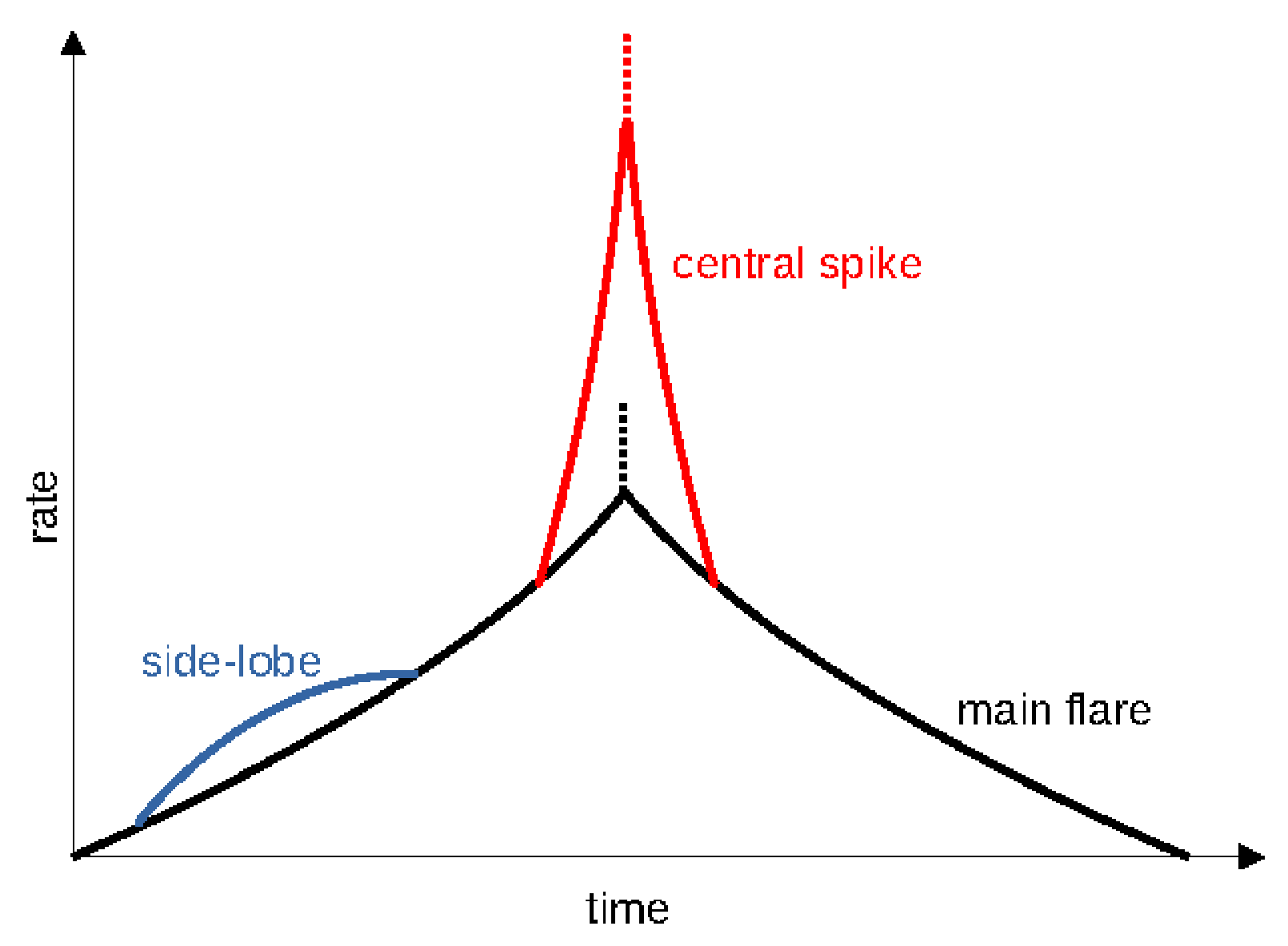}}
\caption{Illustration of AFP substructures. Central spike is seen only in CV MV\,Lyr during the high state. If a fake spike due to Poisson noise appears, it is on the top of the real central spike and confuses the maximal flux (red dotted line). If the real spike is not present, the fake spike can appear on the top of the main flare like in Fig.~\ref{profile_spike_bias} (black doted line).}
\label{flare_ilustration}
\end{figure}

The AFP consists of two (sub)structures; main flare\footnote{The main flare in AGNs is equivalent to the wider base around the central spike in MV\,Lyr during the high state (Fig.~\ref{profiles_mvmodif}).} described by one or two exponential functions in the rise and decay, and a large side-lobe during higher overall flux (Fig.~\ref{flare_ilustration}). We also note that the averaged flare of Cyg\,X-1 has a similar side-lobe structure in the rising phase (\citealt{negoro2001}). Discussion of individual substructures and their physical meaning follows.

\subsection{Optical vs. X-ray AFPs}

Important is to explain first why we compare optical radiation in CVs with X-rays in AGNs. Based on accretion fluctuation propagation scenario every inhomogeneity generated in outer disc regions propagates inwards. All subsequent mass accretion fluctuations during the accretion flow are coupled, and modulates the X-ray radiation liberated in the inner disc. Every fluctuation entering the corona from the inner disc must therefore modulate the optical radiation from the inner disc, and the X-rays liberated in the corona. The similarity of profiles implies that the physical sources of optical light and X-rays are close or correlated.

Another possible explanation is based on reprocessing of X-rays into optical like proposed in the MV Lyr case (\citealt{scaringi2014}). In this case the geometrically thick corona surrounding the geometrically thin disc is radiating in X-rays. These X-rays are reprocessed by the underlying thin disc. In the case of CVs and AGN with truncated discs, the central evaporated accretion flow (corona) can be the source of X-rays which irradiates the inner disc. The latter is reprocessing the X-rays into optical radiation then. Similar flare profiles are expected then.

However, the reprocessing situation is not the same in AGNs and CVs. \citet{alston2019} concluded that UV PDS in IRAS\,13224-3809 is behaving very differently compared to X-rays. This suggests different physical origin of the variability and does not support the reprocessing scenario. Nevertheless, the X-ray variability is generated in the X-ray corona. Contrary, the PDS in optical (\citealt{scaringi2012a}), UV and X-rays (\citealt{dobrotka2017a}) in MV\,Lyr have the same characteristics suggesting the reprocessing. Therefore, the optical variability in MV\,Lyr is an image of the X-ray variability generated by the corona. This suggests that our motivation to compare the two bands in AGNs and in MV\,Lyr is justified.

\subsection{AFPs in AGNs vs. CVs}

While CVs in the high state have discs developed up to WD surface, the quiescent discs are truncated. The latter are more similar to AGNs where the discs are likely truncated too. This implies an explanation why the quiescent AFP in MV\,Lyr is similar to AGNs. Let us suppose that the very central part of the disc is the source of the optical central spike in the AFP of MV\,Lyr. The CVs in the high state have this inner disc, while quiescent CVs and AGNs do not. As a consequence, the central spike disappears in systems with truncated discs.

\begin{figure}
\resizebox{\hsize}{!}{\includegraphics[angle=0]{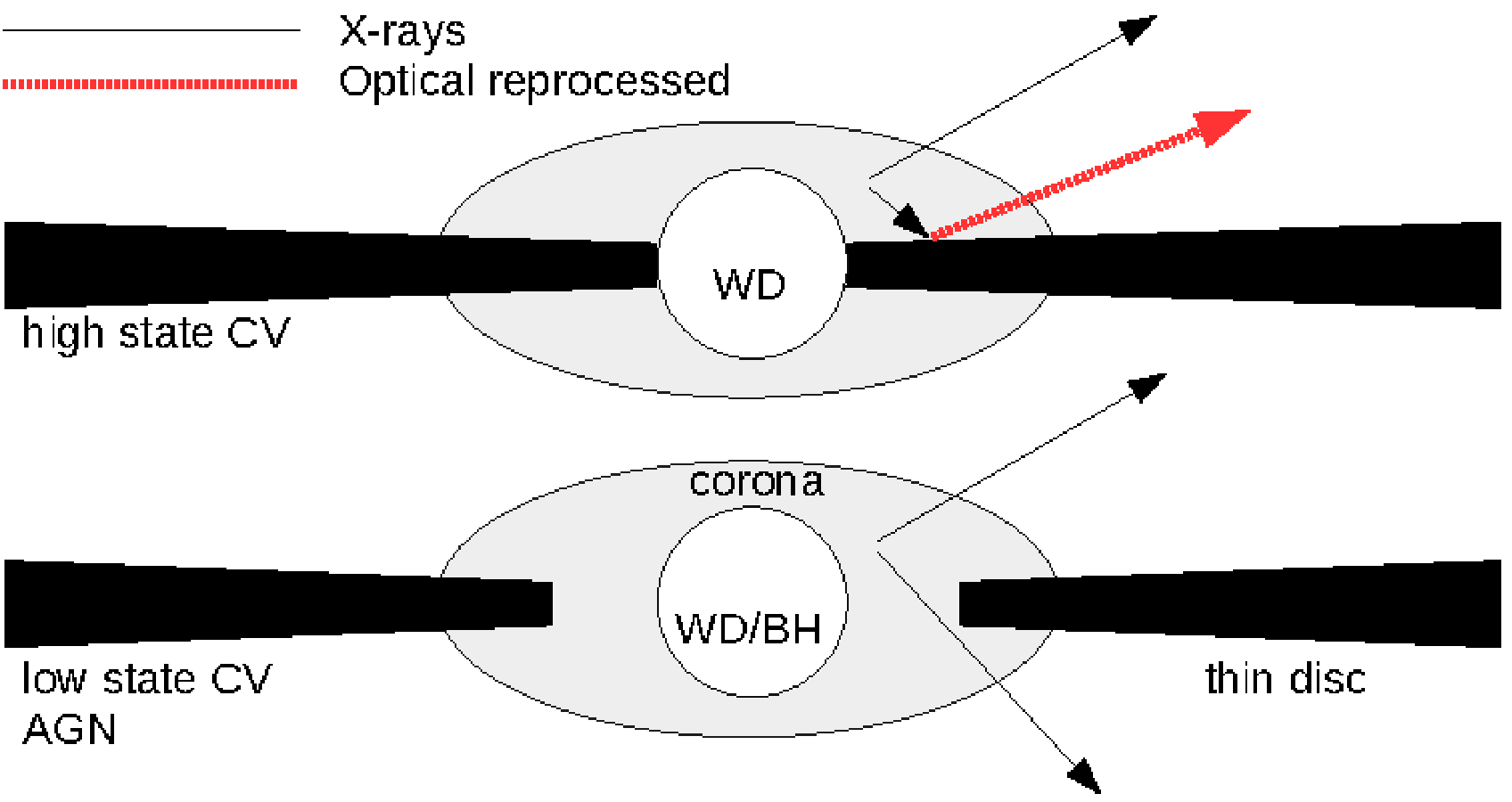}}
\caption{Illustration of the accretion geometry with white dwarf (WD), black hole (BH), geometrically thin and thick (corona) discs. The arrows show the "native" X-ray and reprocessed radiations.}
\label{model}
\end{figure}

The situation is depicted in Fig.~\ref{model}. It is still not sure how the corona looks like. It is some hot plasma near the BH. Many authors use the lamp-post model where the corona is above the BH on the rotation axis (\citealt{martocchia1996}). We investigate the sandwich model where the corona is produced by evaporation (\citealt{meyer1994}) above or below the geometrically thin disc, and forms a geometrically thick disc. In the innermost regions the geometrically thin disc is evaporated completely (truncated) and an advection dominated accretion flow forms (see e.g. \citealt{rozanska2000}). This flow may act as a corona and yields a complex accretion geometry (see e.g. Fig.2 in \citealt{liu2009} or Fig.~9 in \citealt{wilkins2014}). But the principal idea of the inner disc as reprocessing region is also valid for the lamp-post geometry.

Following this scenario some substructures of the flare profile are generated by the truncated disc (main flare), while another (central spike) is added during the high state of MV\,Lyr. Fig.~\ref{profiles_mvmodif} shows this situation, i.e. if we ignore the central spike of the MV\,Lyr AFP in the high state and we rescale it, we can find a solution where the CV profile is similar to those of the AGNs.

As already mentioned, X-ray PDS of IRAS\,13224-3809 shows significant PDS patterns at log($f$/Hz) $\simeq$ -3 represented by a narrow Lorentzian plus a broad component (\citealt{alston2019}). MV\,Lyr shows a rather broad component at log($f$/Hz) $\simeq$ -3 in the corresponding X-ray (\citealt{dobrotka2017a}) and optical PDS (\citealt{scaringi2012a}). Therefore, the broad component in IRAS\,13224-3809 is of our interest (Fig.~9 in \citealt{alston2019}). In the case of MV\,Lyr, the PDS bump structure at log($f$/Hz) $\simeq$ -3 results from sharp peaks (central spike) of the flares with time constants of about 200\,s and the side-lobes at about -640\,s and 740\,s from the flare peak (\citet{dobrotka2019}). Unfortunately, in the case of IRAS\,13224-3809 flare we did not find direct evidence for this frequency due to lacking resolution.

It is very attractive to connect this PDS feature at log($f$/Hz) $\simeq$ -3 to the same frequency component detected in MV\,Lyr during the high state, but absent in the low state. How this fits the corona-inner disc interpretation in the AGN? If the corona is the source of such variability, the PDS pattern should be seen in X-rays also in the AGN with truncated disc, because the corona is present. Only the optical and UV counterpart should be absent due to the missing inner disc as the reprocessing region (see model in Fig.~\ref{model}). This is what we observe in IRAS\,13224-3809, i.e. the log($f$/Hz) $\simeq$ -3 signal is seen in X-rays, but is absent in the UV. In MV\,Lyr in the high state the optical PDS matches the X-ray one, but we miss detailed X-ray observation during the low state. If the corona interpretation is correct, the X-ray PDS of MV\,Lyr in the low state may show the log($f$/Hz) $\simeq$ -3 structure though optical PDS did not show such a structure (\citealt{dobrotka2020}).

\subsection{Side-lobes in IRAS\,13224-3809}

Except the main flare we detected also a large side-lobe during higher overall flux (Fig.~\ref{flare_ilustration}). It appears before the maximum, which suggests that the responsible physical mechanism must be present before the main source of the flares.

The amplitude of this substructure increases with increasing overall flux. The increased flux is probably due to increased mass accretion rate. This process can result from modification of the accretion flow geometry. Let's suppose that the corona has a shape of a sandwich (Fig.~\ref{model}). The physical explanation of such geometrically thick disc is based on evaporation of matter due to inefficient cooling via free-free transitions (\citealt{meyer1994}). The more radially distant regions of the thin disc evaporates partially, while the innermost parts are evaporated completely. During the higher flux period such corona can expand radially outwards because if more matter flows through the geometrically thin disc, more energy is generated in outer disc regions. Therefore, more radially distant regions can evaporate. Similar interpretation concluded \citet{kara2013} based on soft time lags in IRAS\,13224-3809, and \citet{wilkins2014} by modeling the relativistically broadened iron K$\alpha$ fluorescence line in 1H\,0707-495. The authors suppose that the X-ray source is more compact during low flux intervals. However, there is no reason to expect any substructure in the flare profile. Such radially expanded corona would generate mass accretion fluctuations on a longer time-scale. Rather a wider flare would be generated which we did not observe.

A combination of corona as a sandwich and some central regions like lamp-post substructure is a possible speculation. Here the lamp-post is not always necessary, but some powerful energy source which reflects the deep gravitational potential well near the central part of the disk is necessary (see e.g. Fig.~4 in \citealt{zdziarski2021}). Therefore, we refer to it as the very central part of the corona. Based on this scenario, the radially extended sandwich corona can be a weak source of X-rays during the low flux periods. Any side-lobe generated by this corona would be too weak, and only the very central source would generate central structure of the flare strong enough to be detected. If the flux increases, the larger mass accretion rate results in larger particle density $n$ of the corona. Such hot corona radiates via free-free processes where the emissivity is proportional to $n^2$. Any mass accretion fluctuation entering this denser corona becomes observable first as increased radiation from the sandwich corona. Subsequently the fluctuation enters the very central region where it radiates more prominently.

Alternatively, the sandwich corona is too compact and localized only in the very central regions close to the BH during the low flux state. As a consequence the side-lobe blends with the main flare, i.e. the two sub-structures are not distinguishable. During the high luminosity state the sandwich shaped corona is radially more extended allowing the side-lobe to be distinguished from the main flare. This scenario is practically the same as the previous-one. The radially more extended corona is just the case where the emissivity due to larger $n$ is sufficient enough up to larger radii.

Even if the proposed scenario is too speculative, the side-lobe in the AFP suggests that an additional source of X-rays appears during the high flux period. This source should appear earlier in the accretion flow, because it precedes the main flare. The described scenario is an independent confirmation of similar process discovered by \citet{wilkins2014}. The authors studied X-ray observations of Mrk\,335, and described evolution of the corona geometry during flux variability. On long time-scales, they find that during the high flux epochs the corona has expanded, and subsequently contracted to more compact form in the intermediate and low flux epochs.

However, we found that the rise profiles in MV\,Lyr and the AGNs are very similar. If the main flare is associated to the very central substructure like the lamp-post part of the corona, the similarity can imply similar corona geometry also in CVs. It is hard to conclude whether CVs also have such lamp-post substructure of the corona, but it is sure that CVs have boundary layers. Based on the similarity of the AFPs, some very central corona substructure could be the equivalent of the boundary layer in CVs. However, the energetics can be challenging. Since in BH disks, approximately half of the gravitational energy is released in the disk (see e.g. \citealt{pringle1981}), and the rest of energy is lost (disappears into a BH). On the other hand, in CVs and neutron star cases, the rest of energy is expected to be released on the surface of the star or in the boundary layer. Therefore, the energy release in the case of AGN should be less effective, which is hard to compare with MV\,Lyr AFP since our AGN and CV data are not in the same band. Anyhow, in both AGNs and CVs the mass accretion fluctuation first propagates in the sandwiched corona, modulates the local radiation, and subsequently enters the very central substructure in AGNs, or boundary layer in CVs where the main flare is generated.

So far we described the inner accretion flow as a combination of cold geometrically thin disc and a hot geometrically thick corona. However, AGNs show a soft excess suggesting existence of an optically thick warm medium. This medium can be warm corona located in inner parts of the accretion disc (see e.g. \citealt{mehdipour2011,jin2012,rozanska2015,petrucci2018}) and localized just outside the hot optically thin corona (\citealt{kubota2018}). If this warm corona is the speculated structure localized before the source of the main flare, it should be seen mainly in soft X-rays. This is seen in Fig.~\ref{profiles_soft_hard}. Apparently the side-lobe is seen only in soft X-rays. But if we take only the high flux interval, this dominance in soft X-rays disappear, which is puzzling, and complicate the warm corona interpretation.

Finally, the comparison of soft and hard bands in Fig.~\ref{profiles_soft_hard} reminds study of Cyg\,X-1 using ACF which is a symmetric equivalent of the AFP (\citealt{dobrotka2019}). Such ACF function study of Cyg\,X-1 performed \citet{maccarone2000}. The authors found that the ACF is narrower in hard band, and propose drifting blobs in a hot corona as a source of studied fast variability. Such drifting blobs can be an equivalent of the propagating mass accretion fluctuations.

Alternatively, different scenarios can be considered. The side-lobe may be a precursor of the flare, for instance, the magnetic reconnection in an outer part of the accretion disk. \citet{maccarone2000} proposed magnetic flaring as another option for explanation of the observed narrowing of ACF in harder bands.


\subsection{Why typical flare profile?}

As discussed in \citet{dobrotka2019} for MV\,Lyr, an aperiodic mass accretion is most likely to produce the rise part of the flare. The basic idea is based on accretion fluctuation propagation model which is generally accepted as the main source of the flickering in AGNs and CVs. In the model, when the accretion fluctuation reaches the innermost part of the accretion disk, the maximum gravitational energy is released, resulting in an observed relatively sharp peak. Since all accretion fluctuations release energy by the same process and at the same distance from the center, all corresponding flares has the same profile and time scale. As a consequence the width of the flare representing the typical time scale depends on the inner disc radius as shown in Fig.~\ref{model}. Deeper in the potential well is the flare generated, narrower is the flare. This is well seen in MV\,Lyr case where we know well what is the difference in disc between high and low state. In the high state the disc is developed (almost) down to the WD surface, the viscous time scale is short and very narrow central spike is seen. Contrary in the low state, the disc is truncated and the flare is generated further away and has considerably wider shape because of longer viscous time scale. Only variable quantity is the amount of matter in the accretion fluctuation. This affects the quantity of released energy which induces the amplitude.

Finally, the decay of the flare may be produced by the acoustic wave as shown by \citet{manmoto1996}. Moreover, there is an impression that the profiles are very similar on both sides of the flare. It is beyond the scope of this work to investigate this, but following $T_{\rm r}$ and T$_{\rm d}$ parameters in Table~\ref{fit_param} the profiles on each side can have different time scales. The case is variable from object to object and depends on time extension used for fitting. Therefore, the similarity of both sides of the flare is not justified and can be just a coincidence or can have real physical explanation.

\section{Summary and conclusions}

We analysed \xmm\ observations of three selected AGNs. We studied the AFPs. We did the same with \kepler\ data of the CV MV\,Lyr. It appears that all flares have similar shape when the CV is in quiescence. This conclusion is supported by direct fitting of the flares yielding comparable time scales in all four objects. We explain this similarity by the existence of truncated inner disc. Such truncation is not present in high state of the CV, where the disc is developed (almost) up to the WD surface. This supports the reprocessing scenario in which the X-rays are generated by the corona. These X-rays are reprocessed into optical radiation by the underlying geometrically thin disc in the so-called sandwich model. If the disc is truncated, the central disc as the reprocessing region is missing, and the X-ray variability is not seen in optical.

We performed more detailed analysis of AGN IRAS\,13224-3809 because it has the most extensive \xmm\ light curve. \citet{alston2019} found low and high frequency PDS patterns below or close to log($f$/Hz) = -4 and above or close to log($f$/Hz) = -3, respectively. Our direct fitting of the AFP identified only the low frequency component close to log($f$/Hz) = -4.

The IRAS\,13224-3809 AFP shows larger side-lobe on the rising branch. We divided the light curve into low, low-medium and high flux subsamples. The side-lobe does not have the same amplitude in the corresponding AFPs. It is strongest in high flux data, but practically absent in the low flux data. We interpret this by double structured corona having the sandwich part and some very central substructure. Any mass accretion fluctuation enters the sandwich corona and generates the side-lobe. Subsequently it propagates through and reach the very central region where the main flare of the flare is radiated. The sandwich corona is weak during the low flux periods, while any radiated patterns becomes visible only during high flux states.

\section*{Acknowledgement}

AD and PB were supported by the Slovak grant VEGA 1/0576/24. HN was partially supported by Grants-in-Aid for Scientific Research (21K03620) from the Ministry of Education, Culture, Sports, Science and Technology (MEXT) of Japan. We thanks the anonymous referee for very helpful comments concerning the averaged flare profile method.

\bibliographystyle{aa}
\bibliography{mybib}

\label{lastpage}

\end{document}